\documentclass[a4paper,11pt]{article}
\usepackage{pos}

\newcommand{\amu}{\mbox{$a_\mu^{\mathrm{HVP, LO}}$}}

\newcommand{\amuDIS}{$a_\mu^{\mathrm{HVP(LO),DISC}}$}

\usepackage{graphicx}

\newcommand{\fsize}{1.0}

\def\slashchar#1{\setbox0=\hbox{$#1$}           
   \dimen0=\wd0                                 
   \setbox1=\hbox{/} \dimen1=\wd1               
   \ifdim\dimen0>\dimen1                        
      \rlap{\hbox to \dimen0{\hfil/\hfil}}      
      #1                                        
   \else                                        
      \rlap{\hbox to \dimen1{\hfil$#1$\hfil}}   
      /                                         
   \fi}                                         %

\newcommand{\D}{\slashchar{D}}

\title{Progress report on computing the disconnected QCD and the QCD
  plus QED hadronic contributions to the muon’s anomalous magnetic moment}

\author[a,b]{Alexei~Bazavov}
\author[c]{Christine~Davies}
\author[d]{Carleton~DeTar}
\author[e]{Aida El-Khadra}
\author[f]{Steven~Gottlieb}
\author[c]{Dan~Hatton}
\author[f]{Hwancheol~Jeong}
\author[h]{Andreas~Kronfeld}
\author[i]{Peter~Lepage}
\author*[j]{Craig~McNeile}
\author[j]{Gaurav~Ray}
\author[h]{James~Simone}
\author[d]{Alejandro~Vaquero}

\affiliation[a]{Department of Physics and Astronomy, 
Michigan State University,
East Lansing, MI 48824, USA}

\affiliation[b]{Department of Computational Mathematics,
Science and Engineering, 
Michigan State University,
East Lansing, MI 48824, USA}

\affiliation[c]{SUPA, School of Physics and Astronomy, University of
  Glasgow, Glasgow, G12 8QQ, UK}

\affiliation[d]{Department of Physics and Astronomy,
University of Utah, Salt Lake City, UT 84112, USA}

\affiliation[e]{Department of Physics, University of Illinois, Urbana,
  IL 61801, USA}

\affiliation[f]{Department of Physics, Indiana University,
  Bloomington, Indiana, 47405, USA}

\affiliation[h]{Fermi National Accelerator Laboratory, 
Batavia, IL 60510 USA}

\affiliation[i]{Laboratory for Elementary-Particle Physics, Cornell University, Ithaca, NY 14853, USA}

\affiliation[j]{Centre for Mathematical Sciences, University of Plymouth,
UK}

\emailAdd{craig.mcneile@plymouth.ac.uk}

\abstract{%
\vspace*{-6mm}
\textbf{\textsf{Fermilab Lattice, HPQCD, and MILC Collaborations}}\\[1em]

We report progress on calculating the contribution to the anomalous
magnetic moment of the muon from the disconnected hadronic diagrams with light
and strange quarks and the valence QED contribution to the connected diagrams.
The lattice QCD calculations use the highly-improved staggered quark (HISQ)
formulation. The gauge configurations were generated by the MILC Collaboration
with four flavors
of HISQ sea quarks with physical sea-quark masses.
}

\FullConference{%
 The 38th International Symposium on Lattice Field Theory, LATTICE2021
  26th-30th July, 2021
  Zoom/Gather@Massachusetts Institute of Technology
}

\begin{document}
\maketitle

\section{Introduction}

The exciting recent results from the Fermilab Muon g-2  experiment 
for the muon anomalous magnetic moment~\cite{Muong-2:2021ojo},
which are consistent with the previous result from the E821 experiment
at
BNL~\cite{Muong-2:2006rrc},
motivates reducing the errors on lattice QCD calculations of the leading
order hadronic vacuum polarization
contribution to the muon anomalous magnetic moment \amu.
There is a comprehensive recent review~\cite{Aoyama:2020ynm}
of the theoretical calculations
of \amu.

The latest result~\cite{FermilabLattice:2019ugu} 
from our 
program~\cite{Chakraborty:2014mwa,Chakraborty:2015ugp,Chakraborty:2016mwy,FermilabLattice:2017wgj,Chakraborty:2018iyb}
of computing \amu is the value of the 
connected light quark contribution \amu with an error of 1.4\%.
Specifically, the final result for $10^{10}a_{\mu}^{HVP,LO}$=
$699(15)_{u,d}(1)_{s,c,b}$,
included an estimate for the disconnected contribution 
$\Delta a_\mu^{\rho \omega}({\rm disc}) $ = $-5(5) \times 10^{10}$ and 
residual QED corrections of  
$\Delta a_\mu^{\rho \omega}({\rm qed}) $ = $0(5) \times 10^{10}$.
To reduce the final error to below $\approx 0.5$ \% we need to
explicitly calculate the disconnected contributions and the QED
corrections to \amu. In this paper, we report on the progress 
towards this goal.

There are other ongoing projects aimed at reducing the overall error 
on \amu within our collaboration, such as reducing the errors on  the determination
of the lattice spacing~\cite{Lin:2019pia,Hughes:2019ico},
as well as 
reducing the error on the light quark connected 
contribution~\cite{Lahert2021}.

\section{Overview of the analysis method}

The leading-order contribution to the anomalous magnetic moment from
the HVP is 
\begin{equation}
\amu =4 \alpha^2\int_0^\infty dq^2
f(q^2) \hat{\Pi}(q^2)
\end{equation}
where $f(q^2)$ is a weighting factor~\cite{Blum:2002ii}.
$\hat{\Pi}(q^2)$ is the reduced vacuum polarization
\begin{equation}
\hat{\Pi}(q^2) 
= \frac{4 \pi^2}{q^2}
\int_0^\infty dt G(t) 
\left[
q^2 t^2 - 4 \sin^2 (\frac{qt}{2})
\right]
\end{equation}
which can computed from the electromagnetic current correlator
\begin{equation}
G(t) = \frac{1}{3} \int d \textbf{x}
\sum_{f,i} q_f^2 Z_V^2
\langle V_i^f(\textbf{x},t)   V_i^f(\textbf{0},0)
\rangle   
\label{eq:VecCorr}
\end{equation}
where the sum is over flavors $f$ with charges $q_f$
for the vector current $V_i^f$.
We use the kernel 
function method~\cite{Bernecker:2011gh,Feng:2013xsa}:
\begin{equation}
\amu = \sum_{t=0}^{\infty} w_t G(t) 
\label{eq:discExpr}
\end{equation}
to compute $\amu$, where $w_t$ is a known weighting factor.
The correlator $G(t)$ in Eq.~\ref{eq:VecCorr} is very noisy
at large $t$, so the fitted correlator is used after a time $t^\star$.

We use the HISQ action~\cite{Follana:2006rc} for all quark propagators
contributing to $G(t)$ and work on gluon field configurations
that incorporate 2+1+1 flavors of sea quarks using the HISQ 
action~\cite{MILC:2010pul,MILC:2012znn}.

\section{Computatation of the disconnected contribution}

The disconnected contribution requires the non-perturbative calculation of
the quark-line disconnected correlation of vector currents.
For the disconnected contribution we use the taste-singlet vector
operator. The $Z_V$ renormalization factor used was computed
using the RI-SMOM scheme~\cite{Hatton:2019gha}.
We use stochastic random sources to compute the required loops with a
variety of variance reduction 
techniques~\cite{Yamamoto:2018cqm,FermilabLattice:2019dbx}.

The SU(3) structure of the vector current requires that the 
loop for the operator
\begin{equation}
V_j^{ls} = \frac{1}{3} 
(
V_j^{l} - V_j^s
)
\label{eq:su3Vec}
\end{equation}
be computed for the light quark ($l$) and strange quark ($s$).
This object can be efficiently
calculated using 
a technique developed by the ETM
collaboration~\cite{ETM:2008zte}, which was
used in the calculation of the mass of the
$\eta^\prime$~\cite{Jansen:2008wv} and $\omega$
meson~\cite{McNeile:2009mx}, as well as calculations of flavour
singlet nucleon
matrix elements~\cite{Abdel-Rehim:2013wlz}.
This method has also been proposed and further developed by Giusti et
al.~\cite{Giusti:2019kff}.

The measurement in QCD of the loop of the operator in Eq.~\ref{eq:su3Vec}
requires the computation of
\begin{equation}
L(t)^{ls} =  \left\langle \mbox{Tr} 
\left( 
\gamma_\mu \frac{1}{\D + m_l}  - 
 \gamma_\mu \frac{1}{\D + m_s} 
\right) 
\right\rangle
\label{eq:diffdisc}
\end{equation}
where $\D$ is the massless HISQ Dirac operator.

The difference in Eq.~\ref{eq:diffdisc}
can be trivially written down as:
\begin{equation}
L(t)^{ls} =  \left\langle
\mbox{Tr} \left( \gamma_\mu 
 \frac{(m_s - m_l)} {  (\D + m_l)  (\D + m_s)     } 
\right)
\right\rangle
\label{eq:trick}
\end{equation}
The right hand side of Eq.~\ref{eq:trick} can be computed
using noise sources. In
Ref.~\cite{ETM:2008zte} ETM have argued that
Eq.~\ref{eq:trick} has a smaller variance than
Eq.~\ref{eq:diffdisc}. 
\begin{table}
\centering
\begin{tabular}{ c  c  c  c  c  c  }
  \hline             
Ensemble     & a fm   &  $m_\pi$ MeV & L fm & Eigenmodes & $N_{meas}$   \\  \hline
Very coarse  &  0.15  &  134.7       &  4.8   & 300        & 1692  \\
Coarse       &  0.12  &  134.9       &  5.8   &  -         & 787    \\
Fine         &  0.09  &  128.3       &  5.8   & 1000       & 271  \\
  \hline  
\end{tabular}
\caption{HISQ ensembles used in the disconnected analysis} 
\label{tb:disconensemble}
\end{table}
We have computed \amuDIS for the light and strange quarks.
The ensembles used to compute disconnected correlators are listed
in Table~\ref{tb:disconensemble}.
To remove potential subjective bias,
we do a ``blinded analysis.''
The correlators are multiplied by a random 
blinding factor~\cite{Klein:2005di}.
For the fine ensemble,
we used Eq.~\ref{eq:trick} with the 
truncated solver method~\cite{Bali:2009hu,Alexandrou:2012zz}
combined with deflation of 1000 low-modes~\cite{Wilcox:2007ei}.

We broadly follow the analysis method described 
in Ref.~\cite{Chakraborty:2015ugp}. We fit the correlators to a model
of the difference of correlators between flavour singlet and
non-singlet mesons and replace the correlator in 
Eq.~\ref{eq:discExpr} at time larger than
$t^\star$  with the fitted correlator.  

\begin{equation}G(t) = \sum_{i=1}^{N} [
b_i^2 e^{-E_{b_i} t } - a_i^2 e^{-E_{a_i} t }
+
(-1)^t
\left(
c_i^2 e^{-E_{c_i} t } - d_i^2 e^{-E_{d_i} t }
\right)
]
\end{equation}
We use Bayesian fitting~\cite{Lepage:2001ym}
with priors
on the properties of $\rho$, excited $\rho$,
$\omega$ and excited $\omega$.

\begin{figure}
\centering
\includegraphics[scale=\fsize,angle=0]{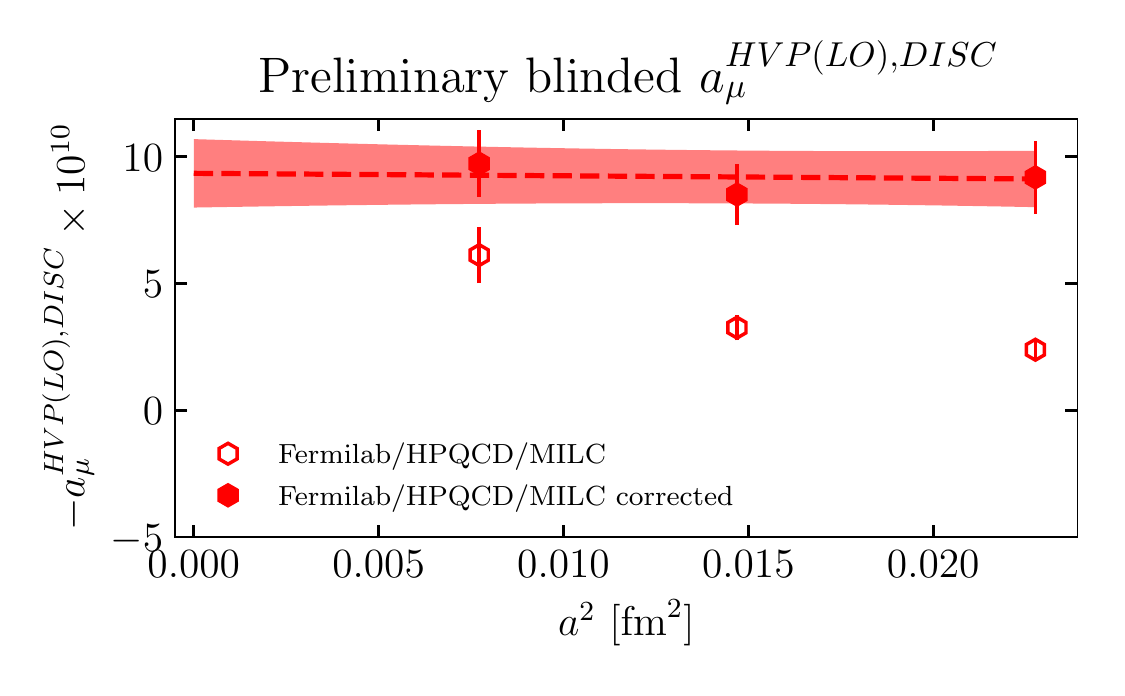}
\caption{\amuDIS as a function of the square of the lattice spacing.}
\label{fg:disc}
\end{figure}

The raw data are corrected by including taste and finite
volume corrections, using
 the model first developed
in Ref.~\cite{Chakraborty:2016mwy} and 
extended in Ref.~\cite{FermilabLattice:2019ugu}.
Figure~\ref{fg:disc} shows the data and the data
corrected for finite volume and taste corrections with
the continuum extrapolation.
\amuDIS is extrapolated to the continuum and physical pion masses
using:
\begin{equation}
\mbox{\amuDIS}(a,m_\pi)  = a_0 \left( 1 + a_1 (a\Lambda)^2 +  a_2  (\frac{m_\pi^2 - m_{\pi,phys}^2} {m_{\pi,phys}^2}
) \right) 
\end{equation}
where $a$ is the lattice spacing in units GeV$^{-1}$,
$\Lambda$ = 0.5 GeV represents the QCD scale.
Broad priors are included for the fit parameters: $a_0$, $a_1$, and $a_2$.


\section{Preliminary window analysis of \amuDIS}

The lattice data can be with the experimental 
$e^+e^- \rightarrow {\rm hadrons} $ scattering 
data~\cite{Bernecker:2011gh,Lehner:2017kuc,Lehner:2020crt}. 
It is useful to compare the $e^+e^- \rightarrow {\rm hadrons}$ scattering
data and lattice data for specific regions in time
using a window function.

For the disconnected correlators there is currently no direct
comparison between experimental data and the lattice correlator,
but it is useful to compare the correlator at different time regions
between lattice calculations,
because this isolates different physics and parts of the correlator
with different statistical properties. The following weight function
in time is used~\cite{RBC:2018dos,Borsanyi:2020mff}
\begin{equation}
W(t; t_1,t_2) \equiv \Theta(t;t_1, \Delta) - \Theta(t;t_2, \Delta)
\label{eq:weight}
\end{equation}
where
\begin{equation}
\Theta(t;t^\prime, \Delta) \equiv \frac{1}{2}
+ \frac{1}{2} \tanh[ \frac{t - t^\prime}{\Delta} ]
\end{equation}
and $\Delta$ = 0.15 fm.

\begin{figure}
\centering
\includegraphics[scale=\fsize]{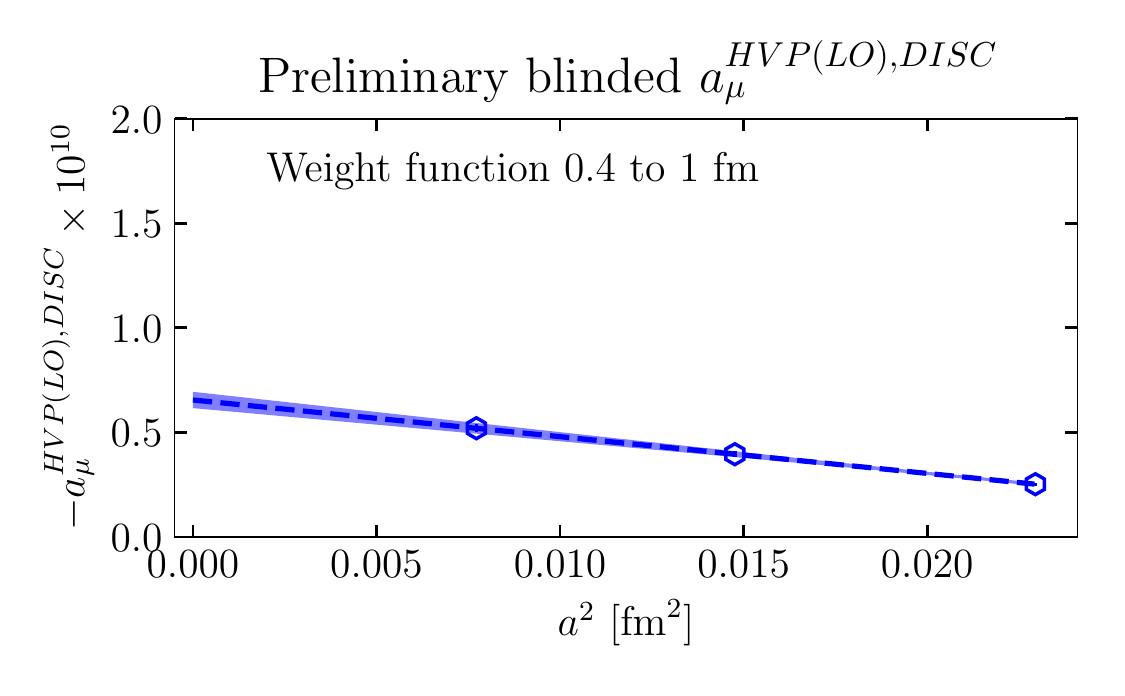}
\caption{The weighted \amuDIS (using the weight function in
Eq.~\ref{eq:weight} with $t_1=0.4$
  fm $t_2=1.0$ fm)
as a function of the square of the lattice spacing.}
\label{fg:discWind}
\end{figure}
The weight function in Eq.~\ref{eq:weight}
with parameters ($t_1=0.4$ fm $t_2=1.0$ fm) is applied to the 
blinded correlators.
\amuDIS is plotted in Fig.~\ref{fg:discWind}
using the specified  weight
function. No taste or finite volume corrections have been applied to
the data in Fig.~\ref{fg:discWind}. 
RBC/UKQCD~\cite{RBC:2018dos}, and BMWc~\cite{Borsanyi:2020mff} also found
a small contribution to \amuDIS from this  window.

\section{Quenched QED corrections to the connected contribution to \amu}

We present preliminary results for the connected QED contribution to
the leading order hadronic contribution to \amu.  We use the
electro-quenched approximation~\cite{Duncan:1996xy,MILC:2018ddw,Hatton:2020qhk} to
partially include the dynamics of QED.  The quenched
QED fields were fixed to the Feynman gauge with zero modes dealt with using
the QED$_{L}$ prescription~\cite{Hayakawa:2008an}.

Other collaborations estimate the QED contribution to QCD calculations by
computing correlators in a perturbative expansion in the electric
charge~\cite{deDivitiis:2013xla}. The RBC/UKQCD collaboration have
compared the perturabtive QED approach to the electroqenched
approach~\cite{Boyle:2017gzv}.  One potential advantage of the perturbative
approach to including QED contributions is that there is a version of
the formalism with
the QED contribution in the infinite volume
limit~\cite{Lehner:2015bga}.

\begin{table}
\centering
\begin{tabular}{ c  c  c  c  c  c  c  c  }
  \hline             
Ensemble & a fm &  $m_\pi$ MeV & L fm  & $N_{meas}$ &
$Z_V^{\mathrm{QCD}}$  
& $Z_V^{\mathrm{QCD+QED}}$  \\  \hline
Very coarse  &  0.15  &  134.7 & 4.8 &  356 & 0.95932(18)  & 0.999544(14)  \\
Coarse (I)     &  0.12  &  132.7 & 5.8  &  300 & 0.97255(22) & 0.999631(24)   \\
Fine   &  0.09  &  128.3 & 5.8  &  128  & 0.98445(11)  & 0.999756(32) \\
  \hline  
\end{tabular}
\caption{The gauge ensembles used to compute QCD + quenched QED
  correlators. The $Z_V$ renormalization factors of the local current 
are from~\cite{Hatton:2019gha}.  }
\label{tb:QEDensemble}
\end{table}

Table~\ref{tb:QEDensemble} lists the ensembles used in the
analysis. The first physical coarse ensemble (I) generated by the
MILC collaboration was used, rather than the coarse
ensemble with the better tuned pion mass in
Table~\ref{tb:disconensemble}, because this ensemble was used to study
quenched QED on the properties of charmonium~\cite{Hatton:2020qhk}.
The quark masses were tuned using only QCD with the lattice
spacing determined from the gradient flow parameter $w_0$ also
determined only including the dynamics of QCD.

The truncated solver method~\cite{Bali:2009hu}, with sloppy inversions
on 16 source times and a precise inversion on one time source per
lattice, was used to compute the vector correlator.
The local vector current was used with the $Z_V$ renormalization
factor determined using the RI-SMOM scheme~\cite{Hatton:2019gha}
in both QCD and QCD+QED (see Table~\ref{tb:QEDensemble}.)

We initially generated correlators at the physical pion mass on the
very coarse ensemble. Unfortunately the results were are noisy with
measurements on 380 lattices. So, following,
BMWc~\cite{Borsanyi:2020mff} we switched to extrapolating from the heavier
light quark masses: $3m_l$, $5 m_l$ and $7 m_l$  to the physical light 
quark mass ($m_l$).

There are
various ways to measure the contribution of the quenched QED on the results,
in this calculation we use the difference.
\begin{equation}
\delta a_\mu^s = a_\mu^s [\mathrm{QCD+qQED}] - a_\mu^s [\mathrm{QCD}]
\end{equation}

We have no estimate of the finite volume 
corrections to $\delta a_\mu^s$; we plan
to quantify these in the future.  This analysis used only neutral
correlators, hence the finite volume corrections from the inclusion of
quenched QED is expected to be much less than for charged correlators.  For
example, HPQCD~\cite{Hatton:2020qhk} found negligible finite volume
corrections for the quenched QED corrections to the mass and decay
constants of the $J/\psi$ and $\eta_c$ mesons, and moments of the
charmed neutral vector correlator.  The analysis of Bijnens et
al.~\cite{Bijnens:2019ejw} found that the finite size effects in the
QED corrections to the hadronic vacuum polarization starts at
$O(\frac{1}{L^3})$ where $L$ is the spatial lattice size.

\begin{figure}
\centering
\includegraphics[scale=\fsize]{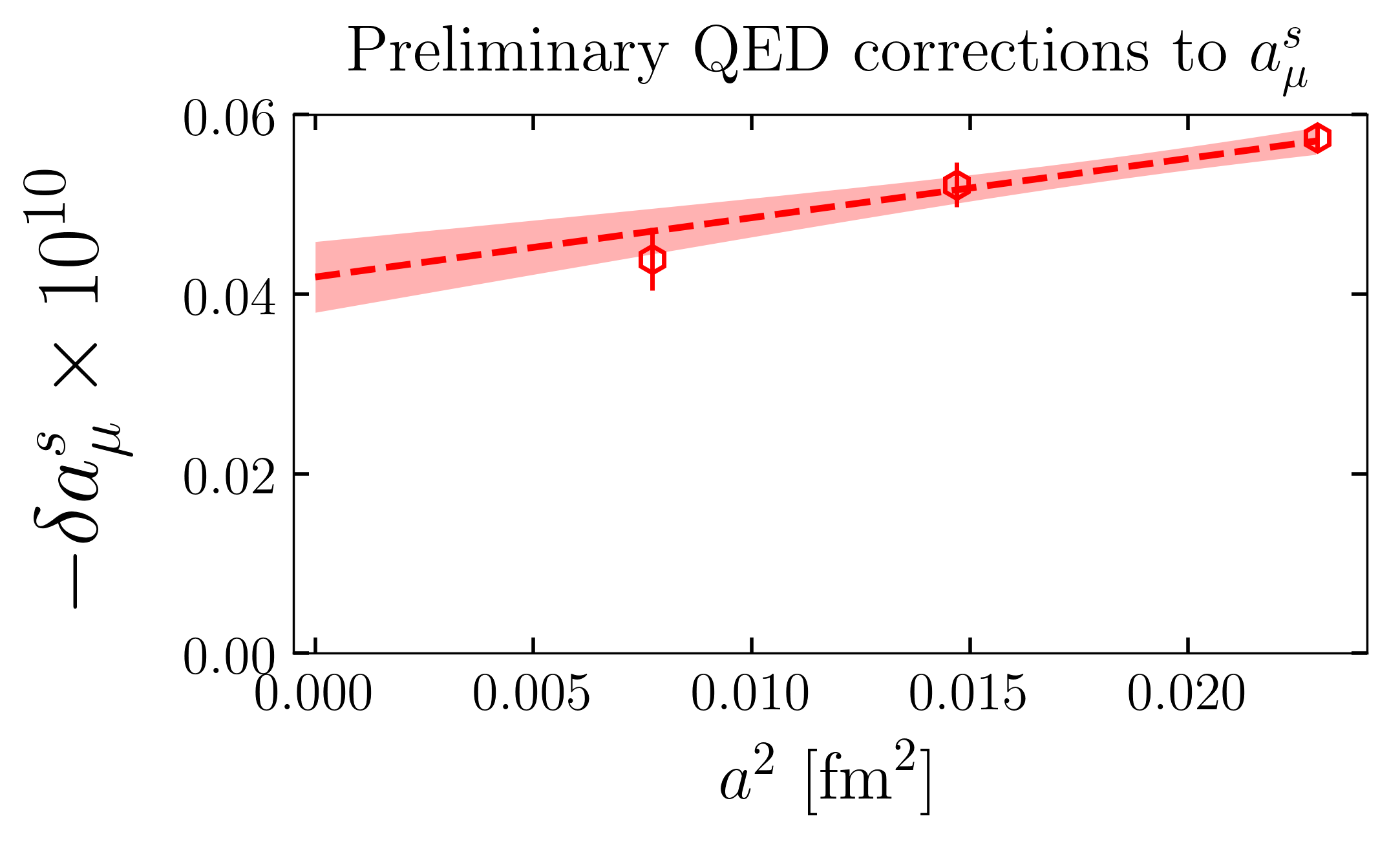}
\caption{Strange quark contribution using lattice spacing and quark
  masses tuned against experiment in QCD.}
\label{fg:amuSQED}
\end{figure}

In Fig.~\ref{fg:amuSQED} we plot the QED contribution to the strange
quark contribution to the HVP
$\delta a_\mu^s$, as a function of the square of the lattice spacing.
To compare our continuum limit result in Fig.~\ref{fg:amuSQED} 
with those reported by BMWc~\cite{Borsanyi:2020mff}, 
ETM collaboration~\cite{Giusti:2019dmu}, and RBC/UKQCD~\cite{RBC:2018dos}, 
one must convert them to a common scheme, which we have not yet done.

\section{Conclusions}

We have presented a progress report on two calculations that aim to decrease the
errors on \amu. We are increasing the statistics on the fine ensemble
for the calculation of the disconnected diagrams. Also we are
increasing the statistics for the valence quenched QED contribution
and investigating the tuning of the light quark masses and lattice spacings
with quenched QED and QCD.

\section*{Acknowledgments}

This work used the Extreme Science and Engineering Discovery
Environment (XSEDE), which is supported by National Science Foundation
grant number ACI-1548562.  Specifically, we uses Ranch and Stampede 2
at the Texas Advanced Computing Center (TACC), located at the
University of Texas at Austin, through XSEDE grant TG-MCA93S002.  This
research used the Cori supercomputer at the National Energy Research
Scientific Computing Center (NERSC), a U.S. Department of Energy
Office of Science User Facility located at Lawrence Berkeley National
Laboratory, operated under Contract No. DE-AC02-05CH11231.
This work used the DiRAC Data Analytic system at the University of
Cambridge, operated by the University of Cambridge High Performance
Computing Service on behalf of the STFC DiRAC HPC Facility
(www.dirac.ac.uk). This equipment was funded by BIS National
E-infrastructure capital grant (ST/K001590/1), STFC capital grants
ST/H008861/1 and ST/H00887X/1, and STFC DiRAC Operations grant
ST/K00333X/1. DiRAC is part of the National E-Infrastructure. We are
grateful to the Cambridge HPC support staff for assistance.

This material is based upon work supported by Science and Technology
Facilities Council, the U.S. Department of Energy, Office of Science,
Office of Nuclear Physics under grant DE-SC0015655 (A.X.K.), and
DE-SC0010120 (S.G.); by the U.S. National Science Foundation under
Grants No. PHY17-19626 and PHY20-13064 (C.D., A.V.); and DOE/NNSA
Exascale Computing Project (17-SC-20-SC) (C.D, A.V., S.G., H.J.).



\begin{thebibliography}{10}

\bibitem{Muong-2:2021ojo}
Muon g-2, B.~Abi {\em et~al.},
\newblock Phys. Rev. Lett. {\bf 126}, 141801 (2021), arXiv:2104.03281.

\bibitem{Muong-2:2006rrc}
Muon g-2, G.~W. Bennett {\em et~al.},
\newblock Phys. Rev. D {\bf 73}, 072003 (2006), arXiv:hep-ex/0602035.

\bibitem{Aoyama:2020ynm}
T.~Aoyama {\em et~al.},
\newblock Phys. Rept. {\bf 887}, 1 (2020), arXiv:2006.04822.

\bibitem{FermilabLattice:2019ugu}
Fermilab Lattice, HPQCD, MILC, C.~T.~H. Davies {\em et~al.},
\newblock Phys. Rev. D {\bf 101}, 034512 (2020), arXiv:1902.04223.

\bibitem{Chakraborty:2014mwa}
HPQCD, B.~Chakraborty {\em et~al.},
\newblock Phys. Rev. D {\bf 89}, 114501 (2014), arXiv:1403.1778.

\bibitem{Chakraborty:2015ugp}
B.~Chakraborty {\em et~al.},
\newblock Phys. Rev. D {\bf 93}, 074509 (2016), arXiv:1512.03270.

\bibitem{Chakraborty:2016mwy}
B.~Chakraborty {\em et~al.},
\newblock Phys. Rev. D {\bf 96}, 034516 (2017), arXiv:1601.03071.

\bibitem{FermilabLattice:2017wgj}
Fermilab Lattice, HPQCD, MILC, B.~Chakraborty {\em et~al.},
\newblock Phys. Rev. Lett. {\bf 120}, 152001 (2018), arXiv:1710.11212.

\bibitem{Chakraborty:2018iyb}
B.~Chakraborty, C.~T.~H. Davies, J.~Koponen, G.~P. Lepage, and R.~S. Van~de
  Water,
\newblock Phys. Rev. D {\bf 98}, 094503 (2018), arXiv:1806.08190.

\bibitem{Lin:2019pia}
Y.~Lin {\em et~al.},
\newblock Phys. Rev. D {\bf 103}, 034501 (2021), arXiv:1911.12256.

\bibitem{Hughes:2019ico}
C.~Hughes, Y.~Lin, and A.~S. Meyer,
\newblock PoS {\bf LATTICE2019}, 057 (2019), arXiv:1912.00028.

\bibitem{Lahert2021}
S.~Lahert {\em et~al.},
\newblock Hadronic vacuum polarization of the muon on 2+1+1-flavor {HISQ}
  ensembles: an update,
\newblock \url{https://indi.to/9GDTd},
\newblock [Talk at lattice 2021; Online; accessed 21-December-2021].

\bibitem{Blum:2002ii}
T.~Blum,
\newblock Phys. Rev. Lett. {\bf 91}, 052001 (2003), arXiv:hep-lat/0212018.

\bibitem{Bernecker:2011gh}
D.~Bernecker and H.~B. Meyer,
\newblock Eur. Phys. J. A {\bf 47}, 148 (2011), arXiv:1107.4388.

\bibitem{Feng:2013xsa}
X.~Feng {\em et~al.},
\newblock Phys. Rev. D {\bf 88}, 034505 (2013), arXiv:1305.5878.

\bibitem{Follana:2006rc}
HPQCD, UKQCD, E.~Follana {\em et~al.},
\newblock Phys. Rev. D {\bf 75}, 054502 (2007), arXiv:hep-lat/0610092.

\bibitem{MILC:2010pul}
MILC, A.~Bazavov {\em et~al.},
\newblock Phys. Rev. D {\bf 82}, 074501 (2010), arXiv:1004.0342.

\bibitem{MILC:2012znn}
MILC, A.~Bazavov {\em et~al.},
\newblock Phys. Rev. D {\bf 87}, 054505 (2013), arXiv:1212.4768.

\bibitem{Hatton:2019gha}
HPQCD, D.~Hatton, C.~T.~H. Davies, G.~P. Lepage, and A.~T. Lytle,
\newblock Phys. Rev. D {\bf 100}, 114513 (2019), arXiv:1909.00756.

\bibitem{Yamamoto:2018cqm}
Fermilab Lattice, HPQCD, MILC, S.~Yamamoto {\em et~al.},
\newblock PoS {\bf LATTICE2018}, 322 (2019), arXiv:1811.06058.

\bibitem{FermilabLattice:2019dbx}
Fermilab Lattice, HPQCD, MILC, C.~E. DeTar {\em et~al.},
\newblock PoS {\bf LATTICE2019}, 070 (2019), arXiv:1912.04382.

\bibitem{ETM:2008zte}
ETM, P.~Boucaud {\em et~al.},
\newblock Comput. Phys. Commun. {\bf 179}, 695 (2008), arXiv:0803.0224.

\bibitem{Jansen:2008wv}
ETM, K.~Jansen, C.~Michael, and C.~Urbach,
\newblock Eur. Phys. J. C {\bf 58}, 261 (2008), arXiv:0804.3871.

\bibitem{McNeile:2009mx}
ETM, C.~McNeile, C.~Michael, and C.~Urbach,
\newblock Phys. Lett. B {\bf 674}, 286 (2009), arXiv:0902.3897.

\bibitem{Abdel-Rehim:2013wlz}
A.~Abdel-Rehim {\em et~al.},
\newblock Phys. Rev. D {\bf 89}, 034501 (2014), arXiv:1310.6339.

\bibitem{Giusti:2019kff}
L.~Giusti, T.~Harris, A.~Nada, and S.~Schaefer,
\newblock Eur. Phys. J. C {\bf 79}, 586 (2019), arXiv:1903.10447.

\bibitem{Klein:2005di}
J.~R. Klein and A.~Roodman,
\newblock Ann. Rev. Nucl. Part. Sci. {\bf 55}, 141 (2005).

\bibitem{Bali:2009hu}
G.~S. Bali, S.~Collins, and A.~Sch{\"a}fer,
\newblock Comput. Phys. Commun. {\bf 181}, 1570 (2010), arXiv:0910.3970.

\bibitem{Alexandrou:2012zz}
C.~Alexandrou, K.~Hadjiyiannakou, G.~Koutsou, A.~O'Cais, and A.~Strelchenko,
\newblock Comput. Phys. Commun. {\bf 183}, 1215 (2012), arXiv:1108.2473.

\bibitem{Wilcox:2007ei}
W.~M. Wilcox,
\newblock PoS {\bf LATTICE2007}, 025 (2007), arXiv:0710.1813.

\bibitem{Lepage:2001ym}
G.~P. Lepage {\em et~al.},
\newblock Nucl. Phys. B Proc. Suppl. {\bf 106}, 12 (2002),
  arXiv:hep-lat/0110175.

\bibitem{Lehner:2017kuc}
RBC, UKQCD, C.~Lehner,
\newblock EPJ Web Conf. {\bf 175}, 01024 (2018), arXiv:1710.06874.

\bibitem{Lehner:2020crt}
C.~Lehner and A.~S. Meyer,
\newblock Phys. Rev. D {\bf 101}, 074515 (2020), arXiv:2003.04177.

\bibitem{RBC:2018dos}
RBC, UKQCD, T.~Blum {\em et~al.},
\newblock Phys. Rev. Lett. {\bf 121}, 022003 (2018), arXiv:1801.07224.

\bibitem{Borsanyi:2020mff}
S.~Borsanyi {\em et~al.},
\newblock Nature {\bf 593}, 51 (2021), arXiv:2002.12347.

\bibitem{Duncan:1996xy}
A.~Duncan, E.~Eichten, and H.~Thacker,
\newblock Phys. Rev. Lett. {\bf 76}, 3894 (1996), arXiv:hep-lat/9602005.

\bibitem{MILC:2018ddw}
MILC, S.~Basak {\em et~al.},
\newblock Phys. Rev. D {\bf 99}, 034503 (2019), arXiv:1807.05556.

\bibitem{Hatton:2020qhk}
HPQCD, D.~Hatton {\em et~al.},
\newblock Phys. Rev. D {\bf 102}, 054511 (2020), arXiv:2005.01845.

\bibitem{Hayakawa:2008an}
M.~Hayakawa and S.~Uno,
\newblock Prog. Theor. Phys. {\bf 120}, 413 (2008), arXiv:0804.2044.

\bibitem{deDivitiis:2013xla}
RM123, G.~M. de~Divitiis {\em et~al.},
\newblock Phys. Rev. D {\bf 87}, 114505 (2013), arXiv:1303.4896.

\bibitem{Boyle:2017gzv}
P.~Boyle {\em et~al.},
\newblock JHEP {\bf 09}, 153 (2017), arXiv:1706.05293.

\bibitem{Lehner:2015bga}
C.~Lehner and T.~Izubuchi,
\newblock PoS {\bf LATTICE2014}, 164 (2015), arXiv:1503.04395.

\bibitem{Bijnens:2019ejw}
J.~Bijnens {\em et~al.},
\newblock Phys. Rev. D {\bf 100}, 014508 (2019), arXiv:1903.10591.

\bibitem{Giusti:2019dmu}
D.~Giusti, V.~Lubicz, G.~Martinelli, F.~Sanfilippo, and S.~Simula,
\newblock PoS {\bf CD2018}, 063 (2019), arXiv:1909.01962.

\end{thebibliography}

\end{document}